\documentclass[lettersize,journal]{IEEEtran}
\usepackage{amsmath,amsfonts}
\usepackage{array}
\usepackage[caption=false,font=normalsize,labelfont=sf,textfont=sf]{subfig}
\usepackage{textcomp}
\usepackage{stfloats}
\usepackage{url}
\usepackage{verbatim}
\usepackage{graphicx}
\usepackage{cite}
\usepackage{booktabs}
\usepackage{xcolor}
\begin{document}

\title{From Intent to Infrastructure: LLM-Driven Agent Compilers for ISAC Networks}

\author{Lijie~Zheng,   
        Xudong~Zhong,
        Baoquan~Ren,
        Xiangwu~Gong,
        Xinghui~Zhu,
        Ji~He,~\IEEEmembership{Member,~IEEE,}
 \thanks{L. Zheng, X. Zhu and J. He are with the School of Computer Science and Technology, Xidian University, Xi'an 710071, China (e-mail: lijzheng@stu.xidian.edu.cn;xhzhu@xidian.edu.cn;garyhej1991@gmail.com).}
 \thanks{X. Zhong, B. Ren and X. Gong are with the Institute of Systems General, Academy of Systems Engineering, Academy of Military Sciences, Beijing 100101, China (e-mail: zxd148367@outlook.com; renbq@126.com;314968641@qq.com).}
}

\markboth{IEEE Communications Magazine}%
{Author \MakeLowercase{\textit{et al.}}: LLM-Driven Agent Compilers for ISAC Networks}

\maketitle

\begin{abstract}
Integrated sensing and communications (ISAC) is moving from proof-of-concept demonstrations to system-level deployment in sixth-generation (6G) networks. Because sensing and communication share hardware, spectrum, and waveform resources, ISAC design now involves many tightly coupled choices, including waveform selection, sensing algorithm setup, resource scheduling, and deployment planning. This design space is already too large to manage well through manual tuning or isolated optimizers. This article introduces the \textit{Agent Compiler}, a large language model (LLM)-enabled compilation layer that translates high-level engineering intent into complete and executable ISAC system configurations. The Agent Compiler works in four stages: intent parsing, task decomposition, policy graph synthesis, and infrastructure mapping. It produces a verifiable intermediate representation called the ISAC Policy Graph (IPG). A runtime engine then deploys the compiled configuration and supports closed-loop adaptation at three levels: fast parameter updates, partial recompilation of affected subgraphs, and full workflow recompilation. The core design principle is strict time-scale separation: the LLM handles slow-loop strategic decisions, while proven algorithms retain real-time control in the fast loop. A UAV-assisted disaster rescue example illustrates the full compilation process. We also discuss open issues, including compilation latency, output reliability, constraint verification, and pipeline security, to guide future research.
\end{abstract}



\section{Introduction}
\IEEEPARstart{I}{ntegrated} sensing and communications (ISAC) is becoming a key part of sixth-generation (6G) wireless networks because it combines data transmission and environmental sensing through shared hardware and spectrum resources~\cite{Liu2022ISAC}. As ISAC moves from laboratory prototypes to larger deployments, however, its design problem becomes much harder. Sensing and communication are now tightly coupled across waveform, antenna, and protocol layers, so designers can no longer treat each part in isolation. Instead, they must handle a joint design space that spans sensing, communication, computation, and control. The interactions across these dimensions are complex and difficult to model, let alone optimize, with conventional methods~\cite{Liu2022ISACLimits}.

In practice, many ISAC systems are still designed through expert rules and repeated manual adjustment. Engineers must choose waveform types, tune sensing chains, allocate time--frequency resources, and decide how nodes should cooperate, all under strong cross-dependencies. For example, changing the waveform ambiguity function affects both target resolution and co-channel interference, while changing beam scheduling influences channel estimation accuracy and radar dwell time at the same time. As networks become denser, more heterogeneous, and more dynamic, such manual design becomes increasingly difficult to scale, especially in emerging applications such as low-altitude operations and urban air mobility~\cite{Lu2024ISACChallenges}.

Existing AI-based methods provide useful improvements, but most of them still solve only part of the full problem. Deep reinforcement learning (DRL) can optimize beamforming, mathematical solvers can handle resource allocation, and model-based methods can support waveform design~\cite{Rezaei2023UAV_ISAC}. Yet these methods usually focus on one subtask at a time and do not connect high-level task requirements to a complete system configuration. Recent studies have also explored large language models (LLMs) for wireless optimization, showing that they can help with decisions such as resource allocation and beam prediction~\cite{Li2026LLM_ISAC}, \cite{Qu2026LLM_MultiAgent}, \cite{Yang2026LLM_SAGSIN}. Even so, in most of these studies the LLM acts as a helper for a local optimization problem rather than as a system-level planner. As a result, the broader problem of compiling task intent into an end-to-end ISAC solution is still open.

In this article, we propose the \textit{Agent Compiler}, an LLM-enabled framework that converts high-level task intent into complete and executable ISAC system solutions. These solutions include waveforms, sensing algorithms, resource strategies, and deployment topologies, together with a closed-loop adaptation mechanism. The Agent Compiler is not intended to replace existing optimization algorithms. Instead, it acts as a \textit{system architect}: it parses intent, breaks a complex task into manageable parts, combines specialized solver modules into a coherent policy graph, and coordinates deployment and adaptation through a runtime loop. In this way, it extends ideas from classical compiler design and LLM-based reasoning agents~\cite{Yao2023ReAct, Schick2024Toolformer} to the real-time and physical constraints of wireless ISAC systems.

The main contributions of this work are summarized as follows:
\begin{itemize}
    \item \textbf{Conceptual Framework.} We introduce the \textit{Agent Compiler} paradigm for ISAC networks, define its role, and clarify how it differs from earlier LLM-based wireless optimization methods.
    \item \textbf{System Architecture.} We present an end-to-end compilation framework that includes intent parsing, task decomposition, policy graph synthesis, physical resource mapping, and runtime adaptation in a closed-loop design.
    \item \textbf{Experimental Validation.} We study a UAV-assisted disaster rescue scenario to show how the proposed framework can generate coherent ISAC configurations under dynamic conditions.
    \item \textbf{Critical Analysis.} We discuss key challenges for LLM-driven compilation in wireless infrastructure, including latency, verification, robustness, and security, and outline directions for future work.
\end{itemize}


\begin{table*}[t]
\centering
\caption{Comparison of ISAC Design Paradigms}
\label{tab:comparison}
\renewcommand{\arraystretch}{1.3}
\begin{tabular}{>{\centering\arraybackslash}p{3.6cm}>{\centering\arraybackslash}p{4.6cm}>{\centering\arraybackslash}p{6.0cm}>{\centering\arraybackslash}p{1.5cm}}
\toprule
\textbf{Approach} & \textbf{Key Capability} & \textbf{Primary Limitation} & \textbf{ISAC Fit} \\
\midrule
Manual rules / expert heuristics
  & Reliable, interpretable, low runtime overhead
  & Rigid; cannot handle dynamic multi-objective trade-offs or unexpected scenarios
  & Weak \\
\midrule
Monolithic optimizer
  & Provably optimal for well-defined instances
  & Fixed model assumptions; limited generalization; weak cross-layer coordination
  & Partial \\
\midrule
DRL / MARL
  & Learns adaptive policies; handles non-convex dynamics
  & High training cost; limited interpretability; weak transfer to new tasks
  & Moderate \\
\midrule
Direct LLM optimization
  & Strong semantic understanding; can decompose tasks without task-specific retraining
  & Too slow for real-time control; limited hardware awareness
  & Weak \\
\midrule
\textbf{Agent Compiler (proposed)}
  & Intent parsing, task decomposition, hardware planning, runtime adaptation
  & Depends on the quality of the solver library; orchestration adds overhead
  & \textbf{Strong} \\
\bottomrule
\end{tabular}
\end{table*}

\section{Why ISAC Needs a New Design Abstraction}

This section explains why ISAC needs an Agent Compiler. We first examine the structural complexity of joint sensing and communication design, then review the limits of current design paradigms, and finally identify the requirements of a new intent-to-infrastructure abstraction layer.

\subsection{The Complexity of ISAC System Design}

ISAC requires radar and communication functions to share waveforms, antennas, spectrum, and computing resources while meeting strongly coupled objectives~\cite{Liu2022ISAC}. This joint operation creates five connected sources of complexity that make traditional design methods increasingly inadequate.

\textbf{Interacting Objectives.} Sensing performance and communication throughput are tied together across protocol layers. A higher radar refresh rate leaves fewer time--frequency resources for data users, while wider beams improve sensing coverage but reduce downlink spatial gain. This coupling runs from waveform design and MAC scheduling to application-level quality of service~\cite{Liu2022ISACLimits}.

\textbf{Different Time Scales.} ISAC systems must react on several time scales at once: millisecond-level beam tracking and adaptive modulation, second-level user or target movement and handover, and minute-level task arrivals and topology changes. These scales are not independent. A new task at the minute scale can immediately create fresh tracking and scheduling requirements at the millisecond scale.

\textbf{Mixed Hardware Platforms.} Modern ISAC deployments may include base stations, edge servers, reconfigurable intelligent surfaces (RIS), and UAVs with very different capabilities~\cite{Yu2023RIS_ISAC}. A single sensing task may need a UAV for angle-of-arrival estimation, a ground station for bistatic cross-checking, and an edge server for data fusion. No single node sees the whole system.

\textbf{Two-Way Sensing--Communication Feedback.} The sensing and communication functions affect each other continuously. Sensing outputs, such as target trajectories and blockage estimates, change beam management and scheduling decisions. In the other direction, channel quality and interference information from the communication side influence which sensing modes are practical~\cite{Lu2024ISACChallenges}.

\textbf{Tight Timing Requirements.} Unlike offline optimization, ISAC often operates under hard latency budgets from milliseconds to seconds. A resource allocation decision that arrives 500\,ms late is not merely suboptimal; in many cases, it is unusable.

\subsection{Limitations of Existing Approaches}

Table~\ref{tab:comparison} compares five design paradigms against these complexity sources. Manual rule-based systems are reliable in well-understood settings, but they adapt poorly when new cross-layer trade-offs appear. Monolithic optimizers can be optimal for a specific problem instance, yet they depend on fixed models and often need to be redesigned when the topology or task changes. DRL and multi-agent RL methods can address dynamic scheduling~\cite{Rezaei2023UAV_ISAC}, but they are expensive to train and often transfer poorly across scenarios. Direct LLM-based optimization adds semantic reasoning, but its latency is fundamentally misaligned with millisecond-level control loops~\cite{Maatouk2024LLM_Telecom}.

\subsection{The Case for an Agent Compiler}

What ISAC needs is a new system layer between high-level task intent and low-level control. This layer should be able to read task intent, break it into subtasks, choose and configure algorithms, map those choices to heterogeneous hardware, and adapt the system when conditions change. This is the role of the \textit{Agent Compiler}.

Importantly, the Agent Compiler does not replace waveform design methods or beamforming algorithms. Instead, it coordinates them by deciding which algorithm to run, with what parameters, on which node, and in what order. The analogy with software compilation is deliberate: a compiler translates a high-level program into executable machine instructions through structured intermediate steps~\cite{Lattner2020MLIR}. In a similar way, the Agent Compiler identifies the waveform family that fits the current task, selects parameters that satisfy the sensing--communication trade-off, and sends the resulting configuration to the appropriate hardware within the required latency budget.


\begin{figure*}[t]
    \centering
    \includegraphics[width=0.95\textwidth]{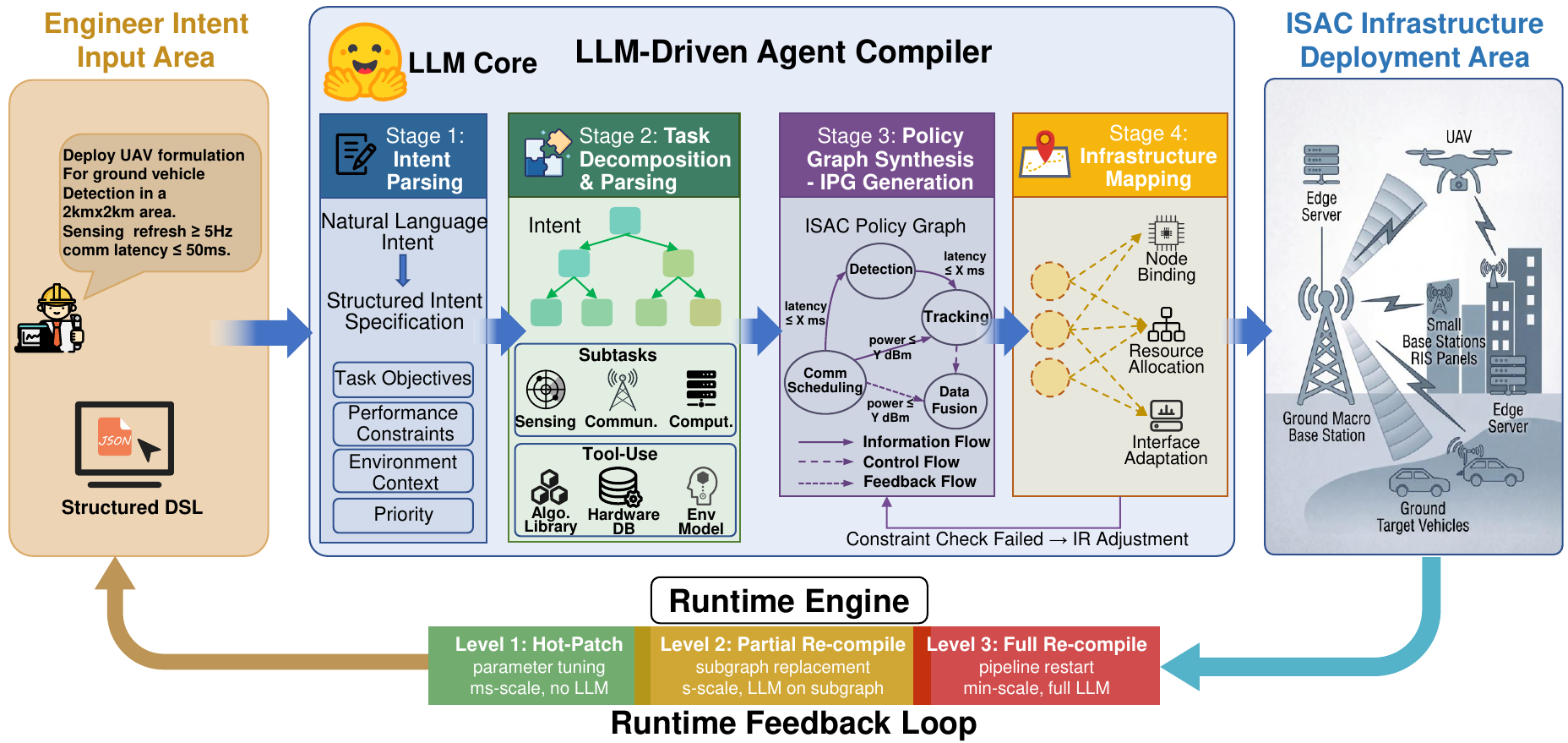}
    \caption{Overall architecture of the LLM-driven Agent Compiler for ISAC networks. Engineer intent is parsed, decomposed, and compiled into an ISAC Policy Graph, which the Runtime Engine deploys and continuously monitors, triggering recompilation at three levels of granularity.}
    \label{fig:overall_arch}
\end{figure*}

\section{The Agent Compiler Architecture}
\label{sec:architecture}

This section describes the proposed Agent Compiler in an end-to-end manner, showing how engineer intent is gradually transformed into executable ISAC configurations through four compilation stages and a runtime adaptation loop.

\subsection{System Overview}

The Agent Compiler is the core of the proposed framework. As shown in Fig.~\ref{fig:overall_arch}, it translates high-level engineer intent into executable ISAC infrastructure configurations. The overall flow is as follows. An engineer first provides a natural-language or structured intent description. The Agent Compiler then processes this input through four internal stages---Intent Parsing, Task Decomposition and Planning, Policy Graph Synthesis, and Infrastructure Mapping---to produce an ISAC Policy Graph (IPG) together with device-level control commands. Finally, the Runtime Engine deploys the configuration, monitors the live ISAC system, and feeds performance information back to the compiler when the system behavior starts to drift away from the original intent.

This architecture is inspired by modern software compiler toolchains~\cite{Lattner2020MLIR}, in which source code passes through several intermediate forms before being lowered to machine instructions. Here, the ``source language'' is human intent, expressed in natural language or a structured command format; the ``intermediate representation'' is a policy graph that captures ISAC task logic; and the ``machine instructions'' are protocol-level control commands sent to heterogeneous radio nodes, UAVs, and edge servers. The LLM serves as the intelligent front end of the compiler. It interprets ambiguous human requests, selects suitable algorithmic building blocks, and reasons about cross-domain dependencies. Deterministic back-end modules then enforce constraints and support real-time execution.

\subsection{Stage 1: Intent Parsing}

The first stage accepts raw input from a network engineer, which may be free-form natural language, a domain-specific language, or a hybrid of the two. Consider the following example: \textit{``Deploy a UAV formation in a 2\,km\,$\times$\,2\,km area for ground vehicle detection, while providing video backhaul for rescue personnel. Sensing refresh rate $\geq$\,5\,Hz, communication latency $\leq$\,50\,ms.''} Working in a tool-augmented reasoning mode~\cite{Yao2023ReAct}, the LLM extracts four types of information from this request: \textit{task objectives} (detection, tracking, and video delivery), \textit{performance constraints} (5\,Hz refresh rate and 50\,ms latency), \textit{environment context} (coverage area, propagation conditions, and mobility), and \textit{priority ordering} (sensing-first, communication-first, or balanced). The result is a structured \textit{Intent Specification} that contains a task list, a constraint dictionary, an environment description, and a priority order. This specification becomes the input to Stage~2.

\subsection{Stage 2: Task Decomposition and Planning}

Stage~2 converts the Intent Specification into executable subtasks across four domains: sensing (operating mode and waveform family), communication (link type and QoS target), computation (on-board, edge, or cloud placement), and control (trajectory and formation). At this stage, the LLM acts as a structured planner~\cite{Huang2024AgentPlanning} and draws from a prepared library of ISAC building blocks exposed as callable tools~\cite{Schick2024Toolformer}.

A key function in this stage is \textit{dependency resolution}. The planner determines the execution order and data flow among subtasks. For example, cooperative localization depends on detection outputs, and communication scheduling must respect spectrum already reserved for sensing. These dependencies are checked against a \textit{hardware capability database tool}, which reports node-level capabilities, and an \textit{environment model tool}, which provides channel statistics and terrain information.

\subsection{Stage 3: Policy Graph Synthesis---IR Generation}

Stage~3 is the most distinctive structural part of the Agent Compiler. Instead of sending device configurations directly, the LLM generates an \textit{ISAC Policy Graph} (IPG), which is a domain-specific intermediate representation inspired by layered intermediate representations in modern compiler systems~\cite{Lattner2020MLIR}. The IPG separates the logical organization of the ISAC task from the hardware details of the target platform, which makes verification, reuse, and incremental modification easier.

\begin{figure}[!t]
    \centering
    \includegraphics[width=0.95\columnwidth]{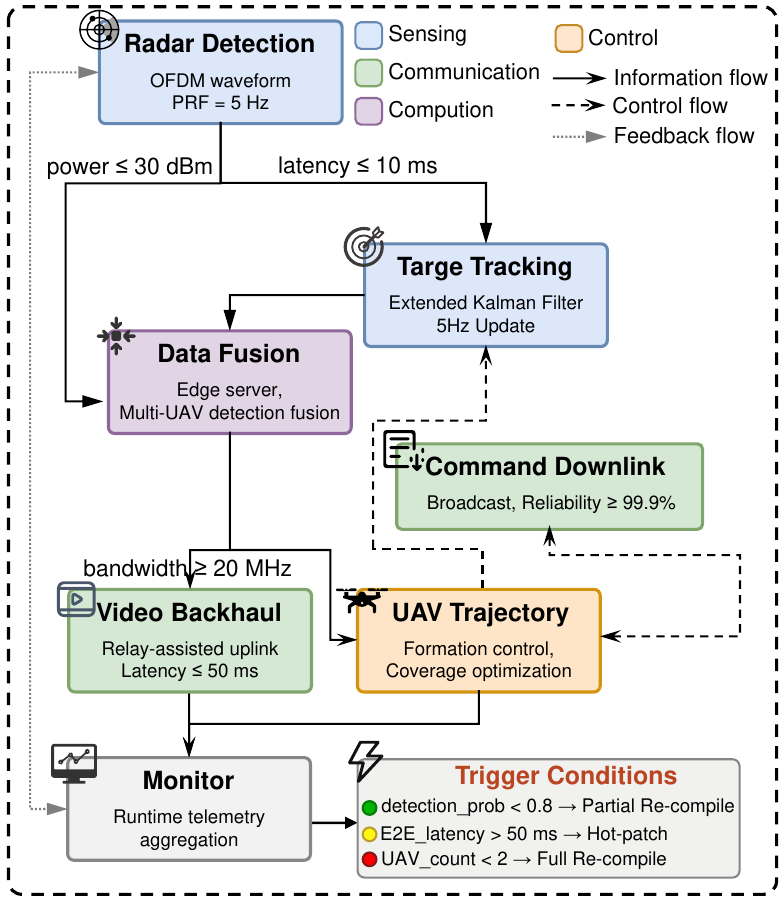}
    \caption{An example ISAC Policy Graph generated for a UAV formation deployment task. Nodes represent sensing, communication, computation, and control subtasks; edges encode information, control, and feedback flows; annotations specify per-node and per-edge operational constraints.}
    \label{fig:ipg_example}
\end{figure}

As shown in Fig.~\ref{fig:ipg_example}, the IPG is a directed attributed graph with three main kinds of elements. \textit{Execution nodes} represent individual subtasks. Each node includes annotations for the chosen algorithm type, parameter ranges (for example, beam dwell time, modulation order, and detection threshold), and execution period. \textit{Directed edges} represent three kinds of flow: information flow edges carry sensing outputs to communication scheduling modules, control flow edges carry decisions to execution nodes, and feedback flow edges return performance monitoring signals to evaluation nodes. In addition, every node and edge carries \textit{constraint annotations}, such as latency budgets, power limits, bandwidth reservations, and minimum accuracy requirements, that can be checked before deployment. Finally, the IPG includes \textit{trigger conditions}, namely logical predicates over monitored metrics that tell the Runtime Engine when recompilation should start.

Using an intermediate representation brings three clear benefits over direct configuration generation. First, the IPG is \textit{verifiable}: its constraint annotations can be checked before any command is sent to hardware, which helps catch infeasible plans early. Second, the IPG is \textit{hardware-agnostic and reusable}, so the same logical graph can be mapped to a fleet of rotary-wing UAVs today and to a mixed fixed-wing and multirotor formation tomorrow without changing the task logic. Third, the IPG supports \textit{incremental recompilation}. If only part of the topology changes, the system can reprocess just the affected subgraph instead of recompiling the entire design.

\subsection{Stage 4: Infrastructure Mapping}

Stage~4 maps the hardware-agnostic IPG into device-executable commands through three steps. First, \textit{node binding} assigns each execution node to a physical device according to capability annotations. Second, \textit{resource allocation} computes an initial spectrum, power, and time-division plan that satisfies edge constraints while keeping some conservative margin. Third, \textit{interface adaptation} translates this plan into device-specific protocol messages, such as waveform parameters, UAV waypoints, and edge offloading commands. Before dispatch, a lightweight constraint checker performs one final feasibility check. If it finds a violation, Stage~4 sends an internal compilation exception back to Stage~3 for targeted subgraph revision, rather than restarting the whole LLM workflow.

\subsection{Runtime Engine: Closed-Loop Adaptation}

After deployment, the Runtime Engine keeps the live system aligned with the engineer's intent through a continuous monitor--evaluate--adapt cycle. The \textit{monitor} module collects sensing metrics, such as detection probability, tracking accuracy, and false alarm rate, together with communication metrics, such as throughput, latency, and packet loss. The \textit{evaluate} module then compares these measurements with the constraint annotations stored in the IPG.

\begin{figure}[!t]
    \centering
    \includegraphics[width=0.95\columnwidth]{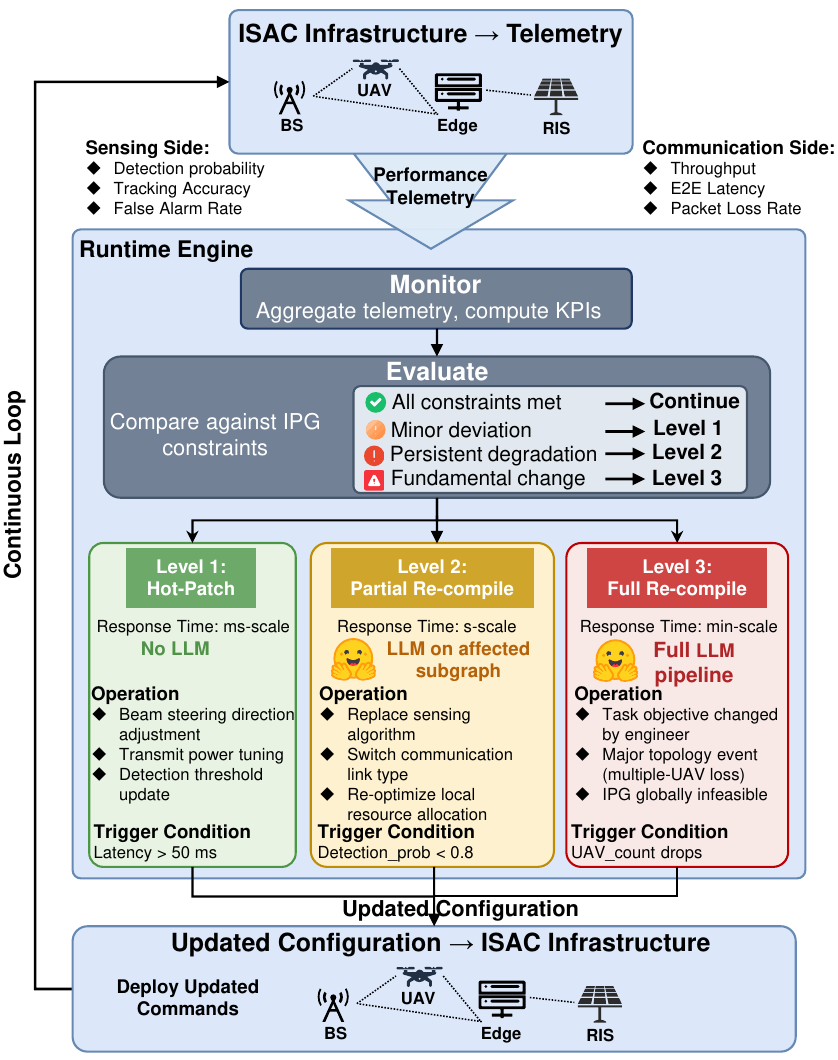}
    \caption{Three-level adaptation architecture of the Runtime Engine. Fast parameter updates operate at millisecond time scales without LLM involvement; partial recompilation engages the LLM for affected subgraphs; full recompilation restarts the workflow.}
    \label{fig:adaptation_levels}
\end{figure}

When a violation appears, the engine chooses one of three adaptation levels, as shown in Fig.~\ref{fig:adaptation_levels}. \textit{Fast parameter updates} handle short-term deviations by adjusting continuous parameters, such as beam direction, transmit power, or detection threshold, within the ranges allowed by the IPG. This level uses deterministic controllers and does not involve the LLM, so it can operate on millisecond time scales. \textit{Partial recompilation} addresses persistent degradation, for example when a sensing algorithm performs poorly in heavy clutter, by calling the LLM only for the affected subgraph on a second-level time scale. \textit{Full recompilation} is reserved for major changes, such as topology disruptions or new mission objectives, and restarts the workflow from Stage~1 or Stage~2 on a minute-level time scale.

The main design principle is a clear separation between \textit{LLM-driven slow-loop decisions} and \textit{algorithm-driven fast-loop control}. The LLM handles strategic decisions on second-to-minute time scales, including intent parsing, algorithm selection, and IPG restructuring. Low-level controllers, such as beamforming, power control, and trajectory tracking, remain on millisecond time scales and do not place the LLM in the critical path~\cite{Maatouk2024LLM_Telecom}.


\section{Case Study}
\label{sec:casestudy}

To show the Agent Compiler workflow end to end, we consider a UAV-assisted disaster rescue scenario in a post-earthquake urban area. Six UAVs are deployed over a 3\,km\,$\times$\,3\,km zone. Four are ISAC-capable sensing UAVs used for survivor detection, and two are relay UAVs used for data forwarding and video backhaul to a remote command center. The engineer submits the following intent: \textit{``Deploy 4 sensing UAVs and 2 relay UAVs for survivor detection in a 3\,km\,$\times$\,3\,km post-earthquake urban area. Maintain a sensing refresh rate $\geq$\,2\,Hz and provide real-time video backhaul with end-to-end latency $\leq$\,100\,ms. Sensing takes priority when resources conflict.''}

\subsection{Compilation Walkthrough}

The Agent Compiler processes this request in four stages. In Stage~1, the LLM converts the input into a structured Intent Specification that contains two task goals (survivor detection and video backhaul), two performance constraints (2\,Hz refresh and 100\,ms latency), an environment description (a 3\,km\,$\times$\,3\,km post-earthquake urban area), an available platform description (four sensing UAVs and two relay UAVs), and an explicit sensing-first priority. In Stage~2, the planner breaks the intent into subtasks across four domains: radar detection and multi-UAV cooperative localization on the sensing side; uplink data aggregation and video relay on the communication side; on-board lightweight inference and edge-based data fusion on the computation side; and UAV trajectory planning on the control side. Dependencies are then resolved. For example, the cooperative localization task requires detection outputs from all four sensing UAVs before it can run. In Stage~3, these subtasks are assembled into an ISAC Policy Graph with explicit constraint annotations, including a sensing--communication time-division ratio of $\tau = 0.7$ that reflects the sensing-first priority, and a trigger condition that starts partial recompilation if the detection probability remains below 0.8 for more than 10 consecutive seconds. In Stage~4, each IPG node is mapped to a physical device, an initial resource plan is computed, and protocol-level commands are dispatched.

\begin{figure}[!t]
    \centering
    \includegraphics[width=0.95\columnwidth]{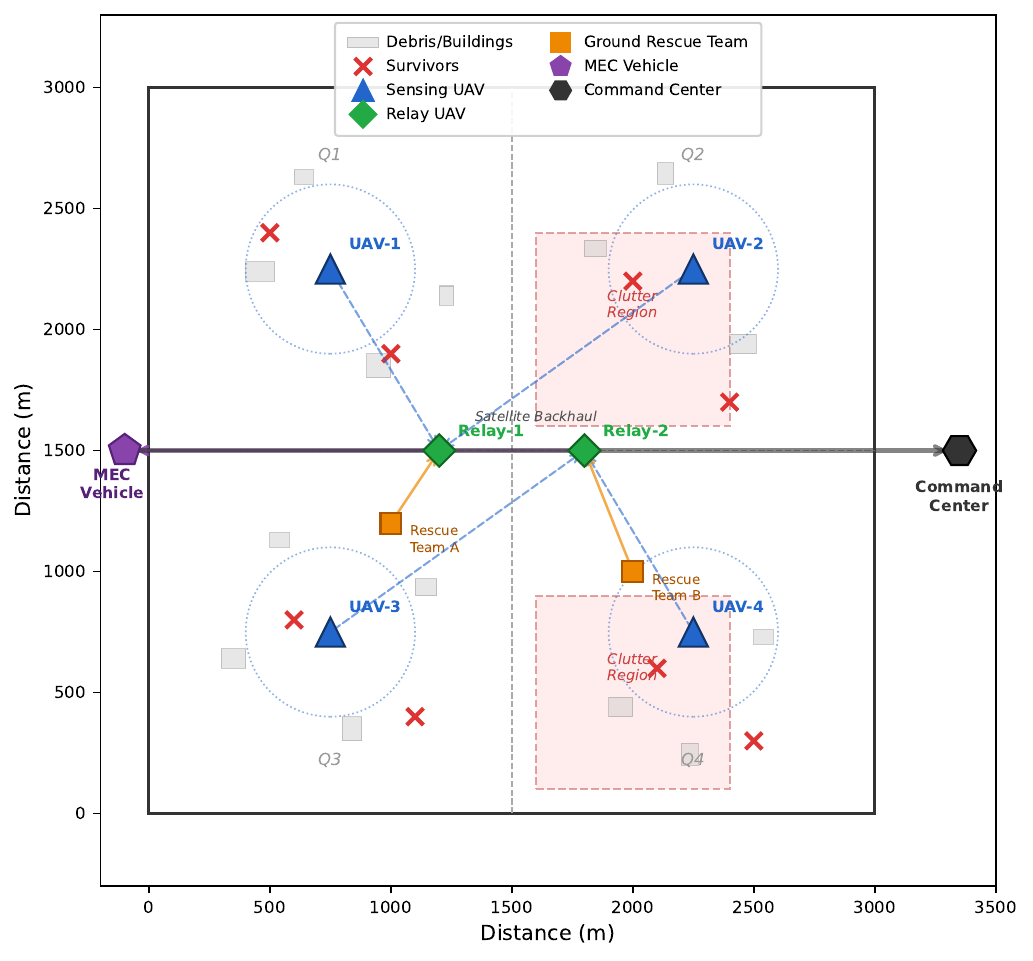}
    \caption{UAV-assisted disaster rescue scenario. Four sensing UAVs cover a 3\,km\,$\times$\,3\,km post-earthquake area and perform ISAC operations together with a mobile edge computing vehicle and a remote command center.}
    \label{fig:scenario}
\end{figure}

Figure~\ref{fig:scenario} shows the resulting deployment. The 3\,km\,$\times$\,3\,km area is divided into four quadrants, each served by one sensing UAV performing dual-function ISAC operations: millimeter-wave radar scanning for survivor detection, shown by the dashed sensing footprints, and uplink data aggregation. Two relay UAVs placed near the quadrant boundaries forward the aggregated sensing data and video streams from ground rescue teams to a mobile edge computing (MEC) vehicle at the edge of the area. The MEC vehicle fuses the data and forwards the results to the remote command center through satellite backhaul. The shaded regions in Q2 and Q4 represent clutter zones caused by metallic debris, where sensing quality is much worse.

\subsection{Runtime Adaptation and Numerical Results}

We evaluate the compiled configuration in a system-level simulation over 300 seconds. The ISAC system uses 500\,MHz bandwidth and a simplified air-to-ground line-of-sight probability channel model. Detection probability is computed from the radar equation as a function of sensing SNR, and communication throughput follows the Shannon capacity bound. At $t = 80$\,s, a strong clutter region caused by metallic debris appears in one quadrant and reduces the local sensing SNR by about 10\,dB. At $t = 200$\,s, one UAV goes offline because of battery depletion, and at the same time a new high-priority search instruction arrives from the command center. We compare five strategies corresponding to the paradigms listed in Table~\ref{tab:comparison}: (i) a \textit{Fixed} baseline with an equal sensing--communication resource split ($\tau = 0.5$) and static UAV trajectories; (ii) a \textit{Rule-based} baseline with predefined if--then adaptation rules; (iii) a \textit{MADDPG} baseline in which a multi-agent deep reinforcement learning policy, trained under nominal channel conditions, outputs per-UAV $\tau$ values at each time step; (iv) a \textit{Direct LLM} baseline that queries an LLM to determine $\tau$ at every decision epoch, with a realistic inference latency of about 3 seconds per call; and (v) the proposed \textit{Agent Compiler} with three-level runtime adaptation.

\begin{figure}[!t]
    \centering
    \includegraphics[width=0.95\columnwidth]{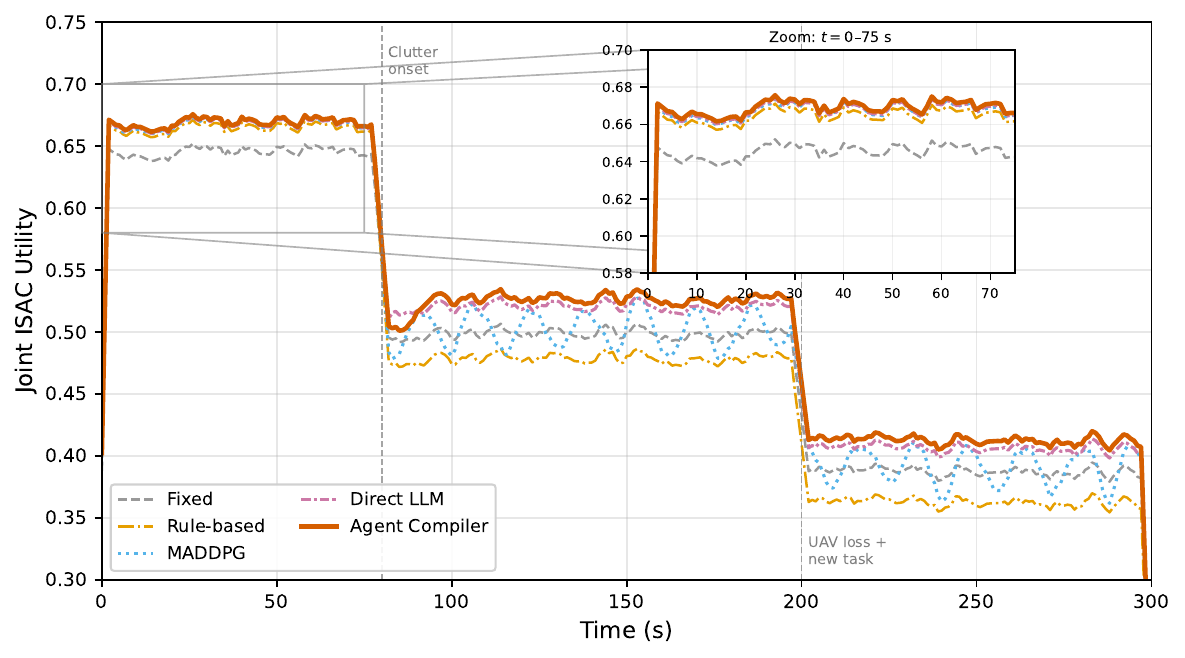}
    \caption{Joint ISAC utility over time for five strategies under dynamic events. Vertical dashed lines mark clutter onset ($t=80$\,s) and UAV loss with a new task directive ($t=200$\,s).}
    \label{fig:results}
\end{figure}

Figure~\ref{fig:results} shows the joint ISAC utility, defined as a weighted combination of normalized detection probability and normalized communication throughput, over time. During the nominal phase ($t < 80$\,s), the zoomed view shows that the five strategies settle at different operating points. The Agent Compiler and MADDPG achieve the highest utility because they adapt $\tau$ to a value close to the optimum of about 0.75, whereas the Fixed baseline with $\tau = 0.5$ performs noticeably worse. After the clutter event at $t = 80$\,s, the performance differences become clear. The Fixed baseline drops immediately and has no recovery mechanism. The Rule-based approach raises the CFAR detection threshold to compensate, but this further reduces detection probability, and its conservative shift of $\tau$ toward sensing makes the problem worse because the degraded sensing channel offers little return. The MADDPG policy was trained under nominal conditions, so once it faces an out-of-distribution state it produces oscillating $\tau$ values and cannot settle near the new optimum. The Direct LLM baseline identifies the correct direction of change and lowers $\tau$ to move more resources to communication, but its 3-second inference delay causes a clear lagged response. By contrast, the Agent Compiler triggers a fast parameter update within one time step, immediately lowering the CFAR threshold and starting to shift $\tau$ toward communication. When the degradation lasts more than 10 seconds, it launches partial recompilation and drives $\tau$ from 0.75 down to about 0.32, recognizing that under severe clutter it is better to rely more on the intact communication link than to keep spending resources on degraded sensing. After the second event at $t = 200$\,s, namely UAV loss together with a new task directive, the Agent Compiler performs full recompilation and redistributes coverage across the remaining three UAVs. Overall, the Agent Compiler achieves the highest average utility of 0.524, followed by Direct LLM (0.519), MADDPG (0.507), Fixed (0.498), and Rule-based (0.487). These results show the practical value of hierarchical, intent-aware runtime adaptation.

\section{Open Challenges and Research Directions}

Despite its promise, the Agent Compiler paradigm still faces several obstacles before it can be deployed in practical ISAC systems. The most important open issues concern runtime efficiency, robustness, verification, privacy, security, and operator trust.

\subsection{Real-Time Performance and Robustness}
Although the Agent Compiler moves most LLM reasoning to the slower compilation stage, the compilation delay can still become a bottleneck when the wireless environment changes suddenly. In highly dynamic ISAC settings, stale plans may quickly lose effectiveness as channel conditions, target motion, or sensing demands shift over time. This challenge is further complicated by the non-deterministic nature of LLMs, since the same intent may lead to different structural outputs across repeated runs. Future work should therefore focus on lightweight models that can run closer to the network edge, pre-compilation caches indexed by intent patterns, and hierarchical execution paths that use validated cached plans for routine updates while invoking full compilation only for genuinely new tasks. In addition, practical systems should include fallback mechanisms that detect abnormal or low-confidence outputs and revert to verified safe-default policies rather than sending risky reconfigurations into the live network~\cite{Chatzistefanidis2024Maestro}.

\subsection{Verifiability, Constraint Satisfaction, and Safe Degradation}
A deployment plan may appear structurally correct while still violating hard physical, regulatory, or system-level constraints, such as interference temperature limits, spectral masks, power budgets, or sensing-quality requirements. For this reason, correctness at the syntax level is not enough for real deployment. Promising directions include static analysis at the IPG level, formal checking over symbolic plan representations, and simulation-in-the-loop validation before execution. It is equally important to design explicit safe-degradation mechanisms. When the requested intent cannot be satisfied under current resource and policy constraints, the compiler should gracefully reduce service quality or narrow the task scope in a transparent manner, instead of silently producing infeasible or unsafe plans.

\subsection{Data Governance, Privacy, and Security}
ISAC systems process large volumes of sensitive data, including radar echoes, Doppler information, environmental maps, and geolocated target tracks. When the Agent Compiler prepares context for the LLM, a key question is how much of this raw information is actually needed. A practical system should minimize unnecessary exposure by replacing full-fidelity sensing data with compact semantic summaries, task-specific abstractions, or privacy-preserving representations whenever possible. Beyond privacy, the compilation pipeline also introduces new security risks. Prompt injection may manipulate the generated plan, corrupted algorithm libraries may distort the compilation outcome, and adversarial physical-layer interference may trigger misleading recompilation behavior~\cite{Zhang2024AdversarialWireless}. Protecting the full pipeline therefore requires joint consideration of data governance, AI safety, and conventional system security.

\subsection{Human Oversight and Trustworthy Operation}
Even a technically strong automation framework will be difficult to deploy if operators cannot understand or control its decisions. Engineers need tools to inspect the compiled IPG, trace how a high-level intent is mapped to specific algorithmic modules, and intervene before execution when necessary~\cite{Huang2024AgentPlanning}. This calls for human-centered interfaces that support explanation, plan review, confidence reporting, and manual override. More broadly, trustworthy operation depends on keeping humans meaningfully involved in the loop, especially in high-stakes wireless scenarios where failures may affect both communication quality and sensing safety. Future research should therefore study not only better compilation accuracy, but also better transparency, accountability, and operator control.


\section{Conclusion}

As ISAC networks grow more complex, isolated point-optimization methods become less able to manage the full system. This article proposed the \textit{Agent Compiler} as a system-level abstraction layer that translates operator intent into structured ISAC Policy Graphs and then instantiates suitable domain-specific modules for real-time execution. The key principle is strict time-scale separation: LLMs handle strategic decisions in the slow loop, while established algorithms retain full authority in the fast loop. More broadly, the same intent-to-infrastructure abstraction may also support the wider 6G vision of integrated communication, sensing, computation, and control.


\bibliographystyle{IEEEtran}
\bibliography{refs}

@article{Liu2022ISAC,
  author={Liu, Fan and others},
  title     = {Integrated Sensing and Communications: Toward Dual-Functional Wireless Networks for {6G} and Beyond},
  journal   = {IEEE J. Sel. Areas Commun.},
  volume    = {40},
  number    = {6},
  pages     = {1728--1767},
  year      = {2022},
}

@article{Liu2022ISACLimits,
  author    = {Liu, An and others},
  title     = {A Survey on Fundamental Limits of Integrated Sensing and Communication},
  journal   = {IEEE Commun. Surveys Tuts.},
  volume    = {24},
  number    = {2},
  pages     = {994--1034},
  year      = {2022},
}

@article{Lu2024ISACChallenges,
  author    = {Lu, Shihang and others},
  title     = {Integrated Sensing and Communications: Recent Advances and Ten Open Challenges},
  journal   = {IEEE Internet Things J.},
  volume    = {11},
  number    = {11},
  pages     = {19094--19120},
  year      = {2024},
}

@article{Li2026LLM_ISAC,
  author    = {Li, Xingwang and others},
  title     = {Recent Advances in Resource Allocation and Beam Prediction for Large Language Models Empowered {ISAC} Systems},
  journal   = {IEEE Commun. Mag.},
  pages     = {1--7},
  year      = {2026},
}

@article{Qu2026LLM_MultiAgent,
  author    = {Qu, Zheyan and Wang, Wenbo and Yu, Zitong and Sun, Boquan and Li, Yang and Zhang, Xing},
  title     = {{LLM} Enabled Multi-Agent System for {6G} Networks: Framework and Method of Dual-Loop Edge-Terminal Collaboration},
  journal   = {IEEE Commun. Mag.},
  pages     = {1--7},
  year      = {2026},
}

@article{Yang2026LLM_SAGSIN,
  author    = {Halvin Yang and Sangarapillai Lambotharan and Mahsa Derakhshani and Lajos Hanzo},
  title     = {{LLM}-Enhanced Space-Air-Ground-Sea Integrated Networks},
  journal   = {IEEE Commun. Mag.},
  pages     = {1--8},
  year      = {2026},
}

@article{Maatouk2024LLM_Telecom,
  author    = {Ali Maatouk and Nicola Piovesan and Fadhel Ayed and Antonio {De Domenico} and M\'{e}rouane Debbah},
  title     = {Large Language Models for Telecom: Forthcoming Impact on the Industry},
  journal   = {IEEE Commun. Mag.},
  volume    = {62},
  number    = {7},
  pages     = {64--69},
  year      = {2024},
}

@inproceedings{Yao2023ReAct,
  author    = {Shunyu Yao and others},
  title     = {{ReAct}: Synergizing Reasoning and Acting in Language Models},
  booktitle = {Proc. Int. Conf. Learn. Representations (ICLR)},
  year      = {2023},
}

@inproceedings{Schick2024Toolformer,
  author    = {Timo Schick and others},
  title     = {Toolformer: Language Models Can Teach Themselves to Use Tools},
  booktitle = {Proc. Advances Neural Inf. Process. Syst. (NeurIPS)},
  volume    = {36},
  year      = {2023},
}

@article{Huang2024AgentPlanning,
  author    = {Xu Huang and others},
  title     = {Understanding the Planning of {LLM} Agents: A Survey},
  journal   = {arXiv preprint arXiv:2402.02716},
  year      = {2024},
}

@inproceedings{Rezaei2023UAV_ISAC,
  author    = {Omid Rezaei and Mohammad Mahdi Naghsh and Seyed Mehdi Karbasi and Mohammad Nayebi},
  title     = {Resource Allocation for {UAV}-Enabled Integrated Sensing and Communication ({ISAC}) via Multi-Objective Optimization},
  booktitle = {Proc. IEEE Int. Conf. Acoust., Speech Signal Process. (ICASSP)},
  pages     = {1--5},
  year      = {2023},
}

@article{Yu2023RIS_ISAC,
  author    = {Zhiyuan Yu and others},
  title     = {Active {RIS}-Aided {ISAC} Systems: Beamforming Design and Performance Analysis},
  journal   = {IEEE Trans. Commun.},
  volume    = {71},
  number    = {12},
  pages     = {7265--7280},
  year      = {2023},
}

@article{Lattner2020MLIR,
  author    = {Chris Lattner and others},
  title     = {{MLIR}: A Compiler Infrastructure for the End of {Moore's} Law},
  journal   = {arXiv preprint arXiv:2002.11054},
  year      = {2021},
}

@article{Chatzistefanidis2024Maestro,
  author    = {Chatzistefanidis, Ilias and Leone, Andrea and Nikaein, Navid},
  title     = {Maestro: {LLM}-Driven Collaborative Automation of Intent-Based {6G} Networks},
  journal   = {IEEE Netw. Lett.},
  volume    = {6},
  number    = {4},
  pages     = {258--262},
  year      = {2024},
}

@article{Zhang2024AdversarialWireless,
  author    = {Wenhan Zhang and Marwan Krunz and Gregory Ditzler},
  title     = {Stealthy Adversarial Attacks on Machine Learning-Based Classifiers of Wireless Signals},
  journal   = {IEEE Trans. Mach. Learn. Commun. Netw.},
  volume    = {2},
  pages     = {276--293},
  year      = {2024},
}

\end{document}